\documentclass[aps,prl,twocolumn,superscriptaddress]{revtex4-1}

\usepackage{graphicx}
\usepackage{dcolumn}
\usepackage{bm}
\usepackage{amssymb}
\usepackage{mathrsfs}
\usepackage{amsmath}
\usepackage{comment}
\usepackage{newtxtext}

\PassOptionsToPackage{breaklinks}{hyperref}
\usepackage{hyperref}
\hypersetup{	colorlinks=true,
			linkcolor=blue,
			citecolor=blue,
			urlcolor=blue,
		}	
\usepackage{url}
\usepackage[hyphenbreaks]{breakurl}
\usepackage{graphicx}

\def\klpionn{K_L \!\to\! \pi^0 \nu \overline{\nu}}
\def\kppipnn{K^+ \!\to\! \pi^+ \nu \overline{\nu}}
\def\klpioxo{K_L \!\to\! \pi^0 X^0}
\def\klpiopiopio{K_L \!\to\! 3\pi^0}
\def\klpipipio{K_L \!\to\! \pi^+ \pi^- \pi^0}
\def\klpiopio{K_L \!\to\! 2\pi^0}
\def\klgg{K_L \!\to\! 2\gamma}

\def\MeV{~{\rm MeV}}
\def\GeV{~{\rm GeV}}

\begin{document}

\title{ Search for $\klpionn$ and $\klpioxo$ Decays at the J-PARC KOTO Experiment }

\newcommand{\InstKorea}{\affiliation{Department of Physics, Korea University, Seoul 02841, Republic of Korea}}
\newcommand{\InstOsaka}{\affiliation{Department of Physics, Osaka University, Toyonaka, Osaka 560-0043, Japan}}
\newcommand{\InstMichigan}{\affiliation{Department of Physics, University of Michigan, Ann Arbor, Michigan 48109, USA}}
\newcommand{\InstChicago}{\affiliation{Enrico Fermi Institute, University of Chicago, Chicago, Illinois 60637, USA}}
\newcommand{\InstNTU}{\affiliation{Department of Physics, National Taiwan University, Taipei, Taiwan 10617, Republic of China}}
\newcommand{\InstArizona}{\affiliation{Department of Physics, Arizona State University, Tempe, Arizona 85287, USA}}
\newcommand{\InstSaga}{\affiliation{Department of Physics, Saga University, Saga 840-8502, Japan}}
\newcommand{\InstKyoto}{\affiliation{Department of Physics, Kyoto University, Kyoto 606-8502, Japan}}
\newcommand{\InstKEK}{\affiliation{Institute of Particle and Nuclear Studies, High Energy Accelerator Research Organization (KEK), Tsukuba, Ibaraki 305-0801, Japan}}
\newcommand{\InstYamagata}{\affiliation{Department of Physics, Yamagata University, Yamagata 990-8560, Japan}}
\newcommand{\InstChonbuk}{\affiliation{Division of Science Education, Chonbuk National University, Jeonju 54896, Republic of Korea}}
\newcommand{\InstJeju}{\affiliation{Department of Physics, Jeju National University, Jeju 63243, Republic of Korea}}
\newcommand{\InstJPARC}{\affiliation{J-PARC Center, Tokai, Ibaraki 319-1195, Japan}}
\newcommand{\InstJINR}{\affiliation{Laboratory of Nuclear Problems, Joint Institute for Nuclear Researches, Dubna, Moscow region 141980, Russia}}
\newcommand{\InstNDA}{\affiliation{Department of Applied Physics, National Defense Academy, Kanagawa 239-8686, Japan}}
\newcommand{\InstOkayama}{\affiliation{Research Institute for Interdisciplinary Science, Okayama University, Okayama 700-8530, Japan}}
\InstKorea
\InstMichigan
\InstNTU
\InstArizona
\InstOsaka
\InstKEK
\InstKyoto
\InstChonbuk
\InstJeju
\InstJPARC
\InstJINR
\InstChicago
\InstNDA
\InstYamagata
\InstOkayama
\InstSaga
\author{J.~K.~Ahn}\InstKorea
\author{B.~Beckford}\InstMichigan
\author{J.~Beechert}\InstMichigan
\author{K.~Bryant}\InstMichigan
\author{M.~Campbell}\InstMichigan
\author{S.~H.~Chen}\InstNTU
\author{J.~Comfort}\InstArizona
\author{K.~Dona}\InstMichigan
\author{N.~Hara}\InstOsaka
\author{H.~Haraguchi}\InstOsaka
\author{Y.~B.~Hsiung}\InstNTU
\author{M.~Hutcheson}\InstMichigan
\author{T.~Inagaki}\InstKEK
\author{I.~Kamiji}\InstKyoto
\author{N.~Kawasaki}\InstKyoto
\author{E.~J.~Kim}\InstChonbuk
\author{J.~L.~Kim}\thanks{Present address: Division of Science Education, Chonbuk National University, Jeonju 54896, Republic of Korea.}\InstKorea
\author{Y.~J.~Kim}\InstJeju
\author{J.~W.~Ko}\InstJeju
\author{T.~K.~Komatsubara}\InstKEK\InstJPARC
\author{K.~Kotera}\InstOsaka
\author{A.~S.~Kurilin}\thanks{Deceased.}\InstJINR
\author{J.~W.~Lee}\thanks{Present address: Department of Physics, Korea University, Seoul 02841, Republic of Korea.}\InstOsaka
\author{G.~Y.~Lim}\InstKEK\InstJPARC
\author{C.~Lin}\InstNTU
\author{Q.~Lin}\InstChicago
\author{Y.~Luo}\InstChicago
\author{J.~Ma}\InstChicago
\author{Y.~Maeda}\thanks{Present address: Kobayashi-Maskawa Institute, Nagoya University, Nagoya 464-8602, Japan.}\InstKyoto
\author{T.~Mari}\InstOsaka
\author{T.~Masuda}\thanks{Present address: Research Institute for Interdisciplinary Science, Okayama University, Okayama 700-8530, Japan.}\InstKyoto
\author{T.~Matsumura}\InstNDA
\author{D.~Mcfarland}\InstArizona
\author{N.~McNeal}\InstMichigan
\author{J.~Micallef}\InstMichigan
\author{K.~Miyazaki}\InstOsaka
\author{R.~Murayama}\thanks{Present address: KEK, Tsukuba, Ibaraki 305-0801, Japan.}\InstOsaka
\author{D.~Naito}\thanks{Present address: KEK, Tsukuba, Ibaraki 305-0801, Japan.}\InstKyoto
\author{K.~Nakagiri}\InstKyoto
\author{H.~Nanjo}\thanks{Present address: Department of Physics, Osaka University, Toyonaka, Osaka 560-0043, Japan.}\InstKyoto
\author{H.~Nishimiya}\InstOsaka
\author{T.~Nomura}\InstKEK\InstJPARC
\author{M.~Ohsugi}\InstOsaka
\author{H.~Okuno}\InstKEK
\author{M.~Sasaki}\InstYamagata
\author{N.~Sasao}\InstOkayama
\author{K.~Sato}\thanks{Present address: Kamioka Observatory, Institute for Cosmic Ray Research, University of Tokyo, Kamioka, Gifu 506-1205, Japan.}\InstOsaka
\author{T.~Sato}\InstKEK
\author{Y.~Sato}\InstOsaka
\author{H.~Schamis}\InstMichigan
\author{S.~Seki}\InstKyoto
\author{N.~Shimizu}\InstOsaka
\author{T.~Shimogawa}\thanks{Present address: KEK, Tsukuba, Ibaraki 305-0801, Japan.}\InstSaga
\author{T.~Shinkawa}\InstNDA
\author{S.~Shinohara}\InstKyoto
\author{K.~Shiomi}\InstKEK\InstJPARC
\author{S.~Su}\InstMichigan
\author{Y.~Sugiyama}\thanks{Present address: KEK, Tsukuba, Ibaraki 305-0801, Japan.}\InstOsaka
\author{S.~Suzuki}\InstSaga
\author{Y.~Tajima}\InstYamagata
\author{M.~Taylor}\InstMichigan
\author{M.~Tecchio}\InstMichigan
\author{M.~Togawa}\thanks{Present address: KEK, Tsukuba, Ibaraki 305-0801, Japan.}\InstOsaka
\author{Y.~C.~Tung}\InstChicago
\author{Y.~W.~Wah}\InstChicago
\author{H.~Watanabe}\InstKEK\InstJPARC
\author{J.~K.~Woo}\InstJeju
\author{T.~Yamanaka}\InstOsaka
\author{H.~Y.~Yoshida}\InstYamagata
\collaboration{KOTO Collaboration} \noaffiliation

\begin{abstract}
	A search for the rare decay $\klpionn$ was performed.
	With the data collected in 2015, corresponding to $2.2 \times 10^{19}$ protons on target, 
	a single event sensitivity of $( 1.30 \pm 0.01_{\rm stat} \pm 0.14_{\rm syst} ) \times 10^{-9}$ was achieved and no candidate events were observed.
	We set an upper limit of $3.0 \times 10^{-9}$ for the branching fraction of $\klpionn$ at the 90\% confidence level (C.L.),
	which improved the previous limit by almost an order of magnitude.
	An upper limit for $\klpioxo$ was also set as $2.4 \times 10^{-9}$ at the 90\% C.L., where $X^0$ is an invisible boson with a mass of $135 \MeV / c^2$.
\end{abstract}

\pacs{	13.20.Eb, 
		11.30.Er,  
		12.15.Hh 
		}

\maketitle

\paragraph*{Introduction.}\hspace{-20pt}
	---The $\klpionn$ decay is a $CP$-violating process and is highly suppressed in the standard model (SM) due to the $s \!\to\! d$ flavor-changing neutral current transition \cite{Littenberg,Kaon_Review}.
	The branching fraction for this decay can be accurately calculated, 
	and is one of the most sensitive probes to search for new physics beyond the SM (see, e.g., Refs.~\cite{Kpinunu_BSM:1,Kpinunu_BSM:2,Kpinunu_BSM:3,Kpinunu_BSM:4,Kpinunu_BSM:5,Kpinunu_BSM:6,Kpinunu_BSM:7,Kpinunu_BSM:8}).
	The SM prediction is $(3.00 \pm 0.30) \times 10^{-11}$ \cite{KLpi0nunuSM}, 
	while the best upper limit was $2.6 \times 10^{-8}$ (90\% C.L.) \cite{PDG2018} set by the KEK E391a experiment \cite{E391a}.
	An indirect upper limit, called the Grossman-Nir bound \cite{GNlimit}, of $1.46 \times 10^{-9}$ is 
	based on the $\kppipnn$ measurement by the BNL E949 experiment \cite{E949}.
	
	The KOTO experiment \cite{KOTOproposal,KOTO} at the Japan Proton Accelerator Research Complex (J-PARC) \cite{J-PARC} is dedicated to studying the $\klpionn$ decay.
	The first physics run was conducted in 2013 and achieved a comparable sensitivity to E391a with 100 h of data taking \cite{KOTO2013}.
	KOTO is also sensitive to the $\klpioxo$ decay \cite{KLpi0X0,KLpi0X0:2}, where $X^0$ is an invisible light boson.
	The upper limit for this decay was set, for the first time in Ref.~\cite{KOTO2013}, as $3.7 \times 10^{-8}$ (90\% C.L.) for the $X^0$ mass of $135 \MeV / c^2$.
	
\paragraph*{Experimental methods and apparatus.}\hspace{-18pt}
	---A 30-GeV proton beam extracted from the J-PARC Main Ring accelerator with a duration of 2 s struck a gold production target \cite{J-PARC_HEF_AuTarget},  
	and secondary neutral particles produced at an angle of $16^\circ$ from the proton beam 
	were transported via the ``KL beam line'' \cite{KOTO_BeamLine} to the experimental area.
	The neutral beam, composed of neutrons, photons, and $K_L$'s, was collimated by two collimators made of iron and tungsten
	to a size of $8 \times 8 \ {\rm cm^2}$ by the end of the 20-m-long beam line.
	The peak $K_L$ momentum was $1.4 \GeV/c$, and the $K_L$ flux was measured \cite{KOTO_KLflux_Shiomi,KOTO_KLflux_Masuda} 
	as $4.2 \times 10^7$ $K_L$'s per $2 \times 10^{14}$ protons on the target at the exit of the beam line.
	The neutron (kinetic energy $>\!100\ {\rm MeV}$) and photon (energy $>\!10\ {\rm MeV}$) fluxes were 
	estimated to be 6 and 7 times larger than the kaon, respectively.
	Neutrons scattered by the collimators outside the nominal solid angle of the beam are referred to as ``halo neutrons.''
	The collimators were aligned with a beam profile monitor \cite{KOTO_BPM} to minimize the halo neutrons.
	
\begin{figure*}
		\includegraphics[width=1\linewidth]{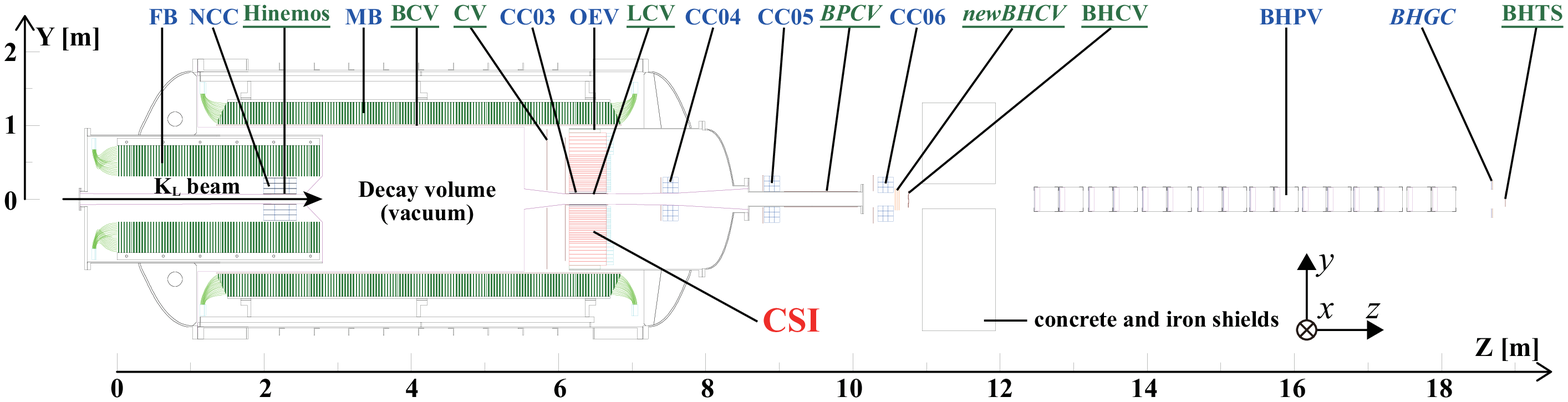}
		\caption{ 	Cross-sectional view of the KOTO detector.
				The beam enters from the left.
				Detector components with their abbreviated names written in blue (in green and underlined) are photon (charged particle) veto counters.
				BPCV, newBHCV, and BHGC are new counters installed after 2013.
				BHCV and BHTS were not used in the analysis.
				}
		\label{fig:KOTODetector}
	\end{figure*}

	A schematic view of the KOTO detector is shown in Fig.~\ref{fig:KOTODetector}.
	The origin of the $z$ axis which lies along the beam direction was the upstream edge of FB, 21.5 m away from the target.
	The $x$ (horizontal) and $y$ (vertical) axes were defined as in the right-handed coordinate system.
	The KOTO detector consisted of the CsI calorimeter (CSI) and hermetic veto counters around the decay volume in vacuum.
	The signature of the $\klpionn$ decay was ``two photons $+$ nothing else'';
	we measured two photons from a $\pi^0$ decay with CSI and ensured that there were no other detectable particles in CSI and veto counters.
	The CSI was composed of 2716 undoped CsI crystals whose length was 50 cm and cross section was 
	$2.5 \times 2.5 \ {\rm cm^2}$ ($5 \times 5 \ {\rm cm^2}$) within (outside) the central $1.2 \times 1.2 \ {\rm m^2}$ region.
	The $15 \times 15 \ {\rm cm^2}$ region at the center of CSI was the beam hole to let the beam particles pass through.
	The veto counters consisted of lead-scintillator sandwich, lead-aerogel, or undoped-CsI counters for photons and plastic scintillators or wire chambers for charged particles.
	The waveform of the signal from all of the detector components was recorded with either 125-MHz digitizers after a Gaussian shaper circuitry \cite{KOTO_125MHzFADC} 
	or 500-MHz digitizers \cite{KOTO_500MHzFADC}.
	Details of the detector components and new components after the 2013 run are explained in Refs.~\cite{KOTO,KOTO2013,KAON2016_Shiomi}.
	
\paragraph*{Data taking.}\hspace{-13pt}
	---This Letter is based on the data set collected in 2015 corresponding to $2.2 \times 10^{19}$ protons on target.
	The power of the primary proton beam increased from 24 to 42 kW during the period.
	The $K_L$ rate at the exit of the beam line was 10~MHz.
	The data acquisition system was triggered by two stages of trigger logic \cite{DAQTrigger_2013run,DAQTrigger_2015run}.
	The first-level trigger (L1) required energy deposition larger than 550 MeV in CSI and 
	the absence of energy deposition in four veto counters which surrounded the decay volume 
	(MB, CV, NCC, and CC03 in Fig.~\ref{fig:KOTODetector}) using loose veto criteria.
	The second-level trigger (L2) calculated the center of energy deposition (COE) in CSI and 
	required the distance from the beam center ($R_{\rm COE}$) to be larger than 165 mm.
	L2 was implemented to reduce the contamination of the $\klpiopiopio$ decay with small missing energy.
	We collected $4.31 \times 10^{9}$ events for the signal sample with these trigger requirements.
	We simultaneously collected samples of $\klpiopiopio$, $\klpiopio$, and $\klgg$ decays for the purpose of normalization and calibration
	by disregarding the L2 decision (and without veto requirements in the L1 decision) with a prescaling factor of 30 (300).

\paragraph*{Reconstruction and event selection.}\hspace{-15pt}
	---The electromagnetic shower generated by a photon in CSI was reconstructed using a cluster of hits in adjacent crystals with energies larger than 3 MeV.
	A $\pi^0$ was reconstructed from two clusters in CSI assuming the $\pi^0 \!\to\! 2\gamma$ decay.
	The opening angle $\theta$ of the two photons was calculated with $\cos \theta = 1 - M^2_{\pi^0}/(2 E_{\gamma_1} E_{\gamma_2})$, 
	where $M_{\pi^0}$ is the nominal $\pi^0$ mass, and $E_{\gamma_1}$ and $E_{\gamma_2}$ are the energies of two photons. 
	The $\pi^0$ decay vertex ($Z_{\rm vtx}$) and transverse momentum ($P_t$) were calculated assuming that the vertex was on the beam axis.
	In the case of the $\klpionn$ decay, the reconstructed $\pi^0$ should have a finite $P_t$ due to the two missing neutrinos.
	Signal candidates were required to have $Z_{\rm vtx}$ in the range of $3000 \!<\! Z_{\rm vtx} \!<\! 4700 \ {\rm mm}$ 
	to avoid $\pi^0$'s generated by halo neutrons hitting detector components.
	The $K_L$ decay probability in the $Z_{\rm vtx}$ range was 3.2\%.
	The candidates were also required to have a $P_t$ in the range of $P_t^{\rm min}\left(Z_{\rm vtx}\right) \!<\! P_t \!<\! 250 \MeV/c$, 
	where $P_t^{\rm min}\left(Z_{\rm vtx}\right)$ was $130 \MeV/c$ in the range of $3000 \!<\! Z_{\rm vtx} \!<\! 4000 \ {\rm mm}$ and
	varied linearly from 130 to $150 \MeV/c$ in the range of $4000 \!<\! Z_{\rm vtx} \!<\! 4700 \ {\rm mm}$.
	This requirement on $P_t$ greatly suppressed events from the $\klpipipio$ decay.

	A series of selection criteria (cuts) based on the energy, timing, and position of the two clusters in CSI were imposed on the candidates.
	We determined all the cuts without examining events inside the region $2900 \!<\! Z_{\rm vtx} \!<\! 5100 \ {\rm mm}$ and $120 \!<\! P_t \!<\! 260 \MeV/c$.
	In order to ensure the consistency with trigger conditions, we required $E_{\gamma_1} + E_{\gamma_2} > 650 \ {\rm MeV}$ and $R_{\rm COE} > 200 \ {\rm mm}$ (trigger-related cuts).
	For each reconstructed photon, we required $100 \!<\! E_{\gamma} \!<\! 2000\ {\rm MeV}$ and 
	the hit position $(x,y)$ to be in the CSI fiducial region of $\sqrt{x^2 + y^2}\!<\!850\ {\rm mm}$ and $\min(|x|,|y|)\!>\!150\ {\rm mm}$ (photon selection cuts).
	The following kinematic cuts on the two photons in CSI were imposed.
	Consistency of the timing of two photons, after taking into account the time of flight from the $\pi^0$ decay vertex to CSI,
	was required to be within 1 ns of each other.
	The distance between the two clusters was required to be larger than $300\ {\rm mm}$ to ensure a clean separation.
	To avoid mismeasurement of photon energies due to three dead channels in CSI, 
	the position of clusters was required to be more than 53 mm apart from those channels.
	The ratio of the energy of two photons, $E_{\gamma_2} / E_{\gamma_1}$ ($E_{\gamma_1} \!>\! E_{\gamma_2}$), was required to be larger than 0.2
	to reduce a class of the $\klpiopio$ background originating from miscombinations of two photons in the $\pi^0$ reconstruction.
	For the same purpose, the product of the energy and the angle between the beam axis and the momentum of a photon was required to be larger than 2500 MeV deg.
	The opening angle of two photons in the $x$-$y$ plane was required to be smaller than $150^\circ$ to reduce the $\klgg$ background, in which the photons are back to back.
	To select $\pi^0$ candidates with plausible kinematics, allowed regions were set on $P_t/P_z - Z_{\rm vtx}$ and $E - Z_{\rm vtx}$ planes,
	where $P_z$ and $E$ are the longitudinal momentum and energy of the $\pi^0$, respectively.
	This cut was effective in reducing the ``CV-$\eta$ background,'' which is described later.
	Events were rejected if there were any hits in the veto counters coincident with the $\pi^0$ decay.
	Cluster-shape and pulse-shape cuts in the CSI (shape-related cuts), defined later, were also imposed on the photons from $\pi^0$ candidates 
	to reduce the background from photon-cluster fusion and neutron showers.

	The signal acceptance $A_{\rm sig}$ was evaluated using \textsc{geant\footnotesize 4}-based \cite{GEANT4:1,GEANT4:2,GEANT4:3} Monte Carlo (MC) simulations.
	Accidental activities in the KOTO detector were taken into account by overlaying random trigger data collected during the data taking.
	The $A_{\rm sig}$ was calculated at 0.52\% after convoluting the reduction from kinematic (57\%), veto (17\%), and shape-related (52\%) cuts.
	The data reduction is summarized in Table~\ref{tab:DataReduction}.

	\begin{table}
		\caption{	Data reduction in each of the selection criteria.
			}
		\label{tab:DataReduction}
		\centering
		\begin{ruledtabular}
		\begin{tabular}{llccc}
			&Selection criteria & & No. events & \\
			\hline
			&Triggered events & & $4.31 \times 10^9$ & \\
			&Two clusters & & $8.74 \times 10^8$ & \\ 
			&Trigger-related cuts & & $2.50 \times 10^8$ & \\ 
			&Photon selection cuts & & $1.75 \times 10^8$ & \\ 
			&Kinematic cuts & & $3.59 \times 10^7$ & \\
			&Veto cuts & & $3.83 \times 10^4$ & \\
			&Shape-related cuts & & 347 & \\
		\end{tabular}
		\end{ruledtabular}
	\end{table}

\begin{figure}
		\begin{center}
			\includegraphics[width=1\linewidth]{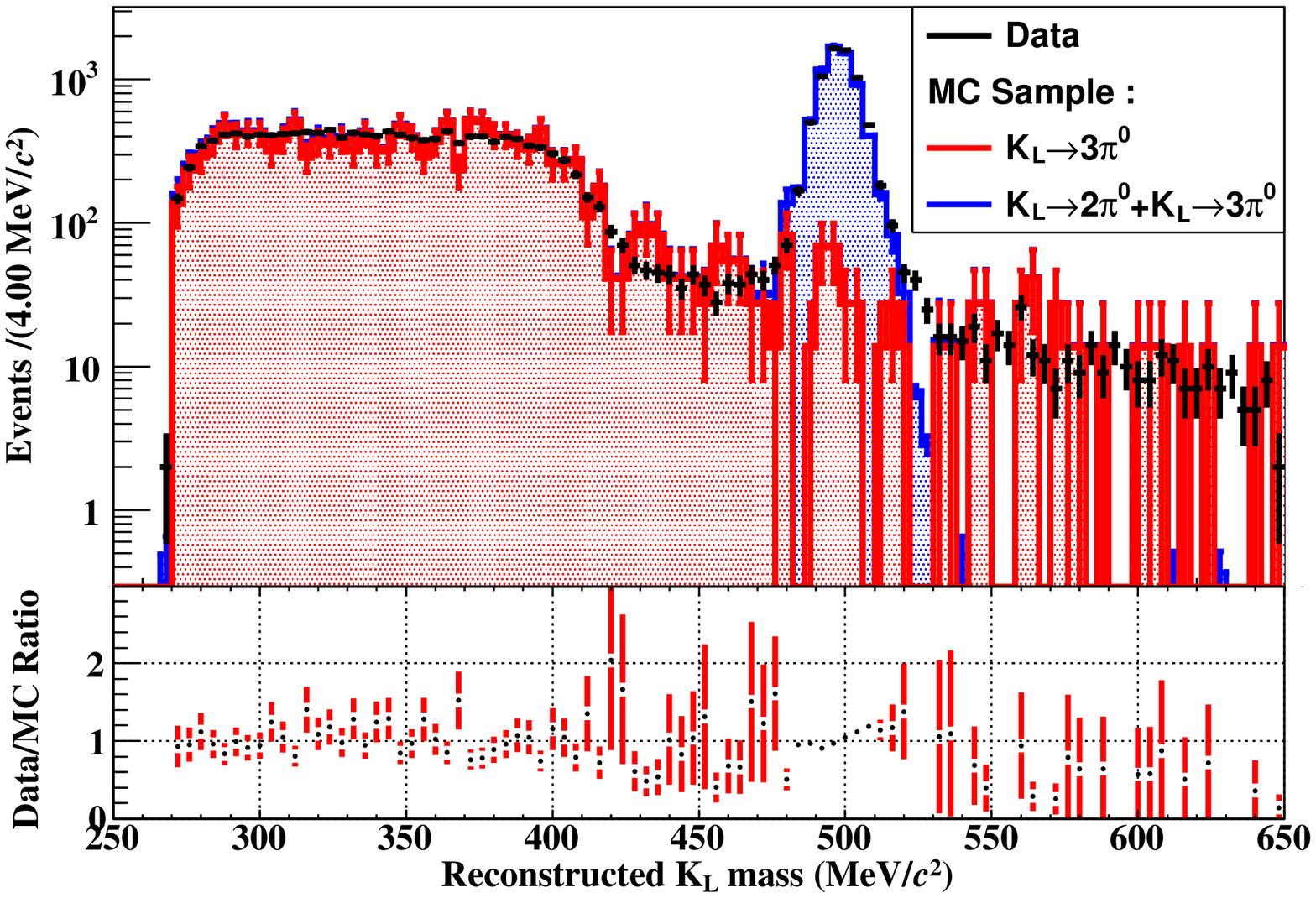}
			\caption{	Four-photon invariant mass distribution of the $\klpiopio$ events after imposing all the cuts except for the $K_L$ mass cut.
					The bottom panel shows the ratio of data and MC (sum of $\klpiopio$ and $\klpiopiopio$) for each histogram bin.
					}
			\label{fig:KLmass}
		\end{center}
	\end{figure}

\paragraph*{Normalization and single event sensitivity (SES).}\hspace{-12pt}
	---The sensitivity for the signal was normalized to the $\klpiopio$ decay;
	events with four photons in CSI were used to reconstruct the $\klpiopio$ events by requiring the pair of $\pi^0$'s 
	with the smallest $Z_{\rm vtx}$ difference among all possible combinations of four photons, together with a series of  kinematic and extra-particle veto cuts. 
	The weighted mean of the two $Z_{\rm vtx}$'s was used to define the decay vertex and select $\klpiopio$ events within the same decay region as the signal.
	Figure~\ref{fig:KLmass} shows the reconstructed $K_L$ mass distribution after imposing all the cuts except for the cut on the $K_L$ mass;
	events within $\pm 15 \MeV/c^2$ around the $K_L$ mass peak were accepted as $\klpiopio$ events.
	
	The single event sensitivity for the $\klpionn$ decay was obtained to be
	\begin{equation}
		{\rm SES} = \frac{1}{A_{\rm sig}} \frac{A_{\rm norm}\ {\rm Br}(\klpiopio)}{p N_{\rm norm}},
		\label{eq:SES}
	\end{equation}
	where $A_{\rm norm}$ is the acceptance for $\klpiopio$ evaluated based on MC simulations,
	${\rm Br}(\klpiopio)$ is the branching fraction of $\klpiopio$ \cite{PDG2018}, 
	$p$ is the prescale factor of 30 used to collect the $\klpiopio$ sample, 
	and $N_{\rm norm}$  is the number of reconstructed $\klpiopio$ events in the data after subtracting the $\klpiopiopio$ contamination.	
	Based on $A_{\rm norm} = 0.36\%$ and $N_{\rm norm} = 1.52 \times 10^4$, 
	the SES was evaluated to be $(1.30 \pm 0.01_{\rm stat} \pm 0.14_{\rm syst}) \times 10^{-9}$.
	The sensitivity is almost an order of magnitude better compared to that of E391a \cite{E391a} and KOTO's first results \cite{KOTO2013}, and comparable to the Grossman-Nir bound.
	The expected number of the SM signal events is 0.023 in this analysis.
	
	The systematic uncertainties in the SES are summarized in Table~\ref{tab:Systematics}.
	The major sources of the uncertainty were the kinematic cuts for the $\klpionn$ selection, the shape-related cuts, 
	and the consistency among the normalization decays $\klpiopio$, $\klpiopiopio$, and $\klgg$.
	The former two were evaluated as follows.
	A sample of $\pi^0$'s from the reconstructed $\klpiopio$ events was used as a validation sample.
	The discrepancy between data and MC acceptance, defined as $(A^i_{\rm MC} - A^i_{\rm data})/A^i_{\rm data}$, 
	where $A^i_{\rm data(MC)}$ represents the acceptance of $i {\rm th}$ cut for data (MC), was used to estimate the systematic uncertainty of the $i {\rm th}$ cut. 
	The sum in quadrature of the uncertainties for each of the kinematic cuts and shape-related cuts 
	resulted in a total systematic uncertainty of 5.1\% for both sets, as shown in Table~\ref{tab:Systematics}. 
	The sensitivity was measured with the $\klpiopiopio$ and $\klgg$ decays,
	and their difference contributed the single largest source of systematic uncertainties of 5.6\%.

	\begin{table}
		\caption{Summary of relative systematic uncertainties in the single event sensitivity.}
		\label{tab:Systematics}
		\centering
		\begin{ruledtabular}
		\begin{tabular}{lD{.}{.}{-1}}
			Source & \multicolumn{1}{c} ~Uncertainty [\%]\\
			\hline
			Trigger effect & 1.9\\
			Photon selection cuts & 0.81\\
			Kinematic cuts for $\klpionn$ & 5.1\\
			Veto cuts & 3.7\\
			Shape-related cuts & 5.1\\
			$K_L$ momentum spectrum & 1.1\\
			Kinematic cuts for $\klpiopio$ & 2.7\\
			$\klpiopio$ branching fraction & 0.69\\
			Normalization modes inconsistency & 5.6\\
			Total &  11\\
		\end{tabular}
		\end{ruledtabular}
	\end{table}

\paragraph*{Background estimation.}\hspace{-15pt}
	\begin{table}
		\caption{Summary of background estimation.}
		\label{tab:BGSummary}
		\centering
		\begin{ruledtabular}
		\begin{tabular}{llc}
			Source & & No. events\\
			\hline
			$K_L$ decay & $\klpipipio$ & 0.05 $\pm$ 0.02 \\
			 & $\klpiopio$ & 0.02 $\pm$ 0.02 \\
			 & Other $K_L$ decays & 0.03 $\pm$ 0.01 \\
			Neutron induced \ \ \  & Hadron cluster & 0.24 $\pm$ 0.17\\
			 & Upstream $\pi^0$ & 0.04 $\pm$ 0.03\\
			 & CV $\eta$ & 0.04 $\pm$ 0.02\\
			Total & & 0.42 $\pm$ 0.18\\
		\end{tabular}
		\end{ruledtabular}
	\end{table}
	---Table~\ref{tab:BGSummary} summarizes the background estimation.
	The total number of estimated background events in the signal region was $0.42 \pm 0.18$.
	We categorized background sources into two groups: $K_L$ decay background and neutron-induced background.

	The $K_L$ decay background was estimated using MC simulations.
	The $\klpipipio$ background was due to the absorption of charged pions in the uninstrumented material downstream of CSI.
	The background from $K_L$ decays was small compared to the neutron-induced background in this analysis.

	The neutron-induced background, which was caused by halo neutrons hitting a detector component, was subdivided into the following three categories.

        The background called ``hadron cluster'' \cite{KAON2016_Nakagiri} was caused by a halo neutron directly hitting CSI and 
        creating a hadronic shower and by a neutron produced in the primary shower to create a second, separated hadronic shower.
	These two showers mimicked the clusters from $\pi^0 \!\to\! 2\gamma$.
	A data-driven approach was taken to estimate this background.
	A control sample was collected in special runs with a 10-mm-thick aluminum plate inserted to the beam core at $Z = -634\ {\rm mm}$ to scatter neutrons.
	Two-cluster events were selected in this control sample with selection criteria similar to those used for the signal sample. 
	Two types of cuts were used to reduce the contamination from these neutron-induced events 
	based on cluster-shape discrimination \cite{PhD_Sato} and pulse-shape discrimination \cite{PhD_Sugiyama}.
	A photonlike cluster was selected by considering several variables based on an electromagnetic shower library produced by the MC simulation. 
	The variable with the most discriminating power between photon and neutron clusters was an energy-based likelihood calculated using the accumulated
	energy distribution in each crystal as a probability density function.
	Additional variables, such as global energy and cluster timing information, were used in minimum chi-square estimations and 
	combined with the energy-based likelihood as inputs to a neural network \cite{TMVA} with a single output variable able to distinguish between electromagnetic and hadronic cluster hypotheses.
	The pulse-shape discrimination used the waveform of readout signal from each CSI crystal.
	The waveform was fitted to the following asymmetric Gaussian:
	\begin{equation}
		A(t) = |A| \exp\left( -\frac{(t-t_0)^2}{2\sigma(t)^2}  \right), 
	\end{equation}
	where $\sigma (t) = \sigma_0 + a(t-t_0)$ depends on the timing difference from the mean of the Gaussian ($t_0$).
	Using templates of the fit parameters, $\sigma_0$ and $a$, obtained in a hadron-cluster control sample and by a photon sample from $\klpiopiopio$,
	a likelihood ratio was calculated to determine whether the clusters are more likely to be the hadron clusters or two photon clusters.
	We evaluated the rejection power of cuts based on these two discrimination variables for the Al-plate control sample by taking their correlation into account. 
	The number of background events was normalized by comparing the numbers of events of the signal sample and of the control sample outside the signal region
	before imposing these cuts, and was estimated to be 0.24.
	Note that this is an overestimate due to kaon contamination in the control sample, 
	which we were unable to subtract quantitatively from the estimation because of the limited statistics.
	
	The background called ``upstream $\pi^0$'' was caused by halo neutrons hitting the NCC counter 
	in the upstream end of the decay volume and producing $\pi^0$'s.
	The reconstructed $Z_{\rm vtx}$ for such decays is shifted downstream into the signal region 
	if the energies of photons are mismeasured to be smaller due to photo-nuclear interactions in CSI,
	or if one photon in the CSI is paired to a secondary neutron interacting in the CSI to reconstruct the $\pi^0$.
	This background was evaluated by simulation, and the yield was normalized to the number of events in the upstream region in the data and MC calculations.
	We estimated the number of this background to be 0.04.

	The background called ``CV $\eta$'' stemmed from the $\eta$ production in the halo-neutron interaction with CV \cite{KOTOdet_CV},
        which was a veto counter of plastic scintillator for charged particles located in front of CSI.
	In this background, when a halo neutron hit CV and produced an $\eta$ meson, and the two photons from the $\eta$ decay hit CSI,
	the two clusters were reconstructed using the $\pi^0$ mass hypothesis which pushes the reconstructed $Z_{\rm vtx}$ upstream into the signal region.
	This background was suppressed by imposing a cut which evaluates the consistency of the shape of the clusters with the incident angle of the photons 
	originated from the $\eta \!\to\! 2\gamma$ decay produced at CV.
	The number of the background events was estimated to be 0.04.
	\begin{figure}
		\begin{center}
			\includegraphics[width=1\linewidth]{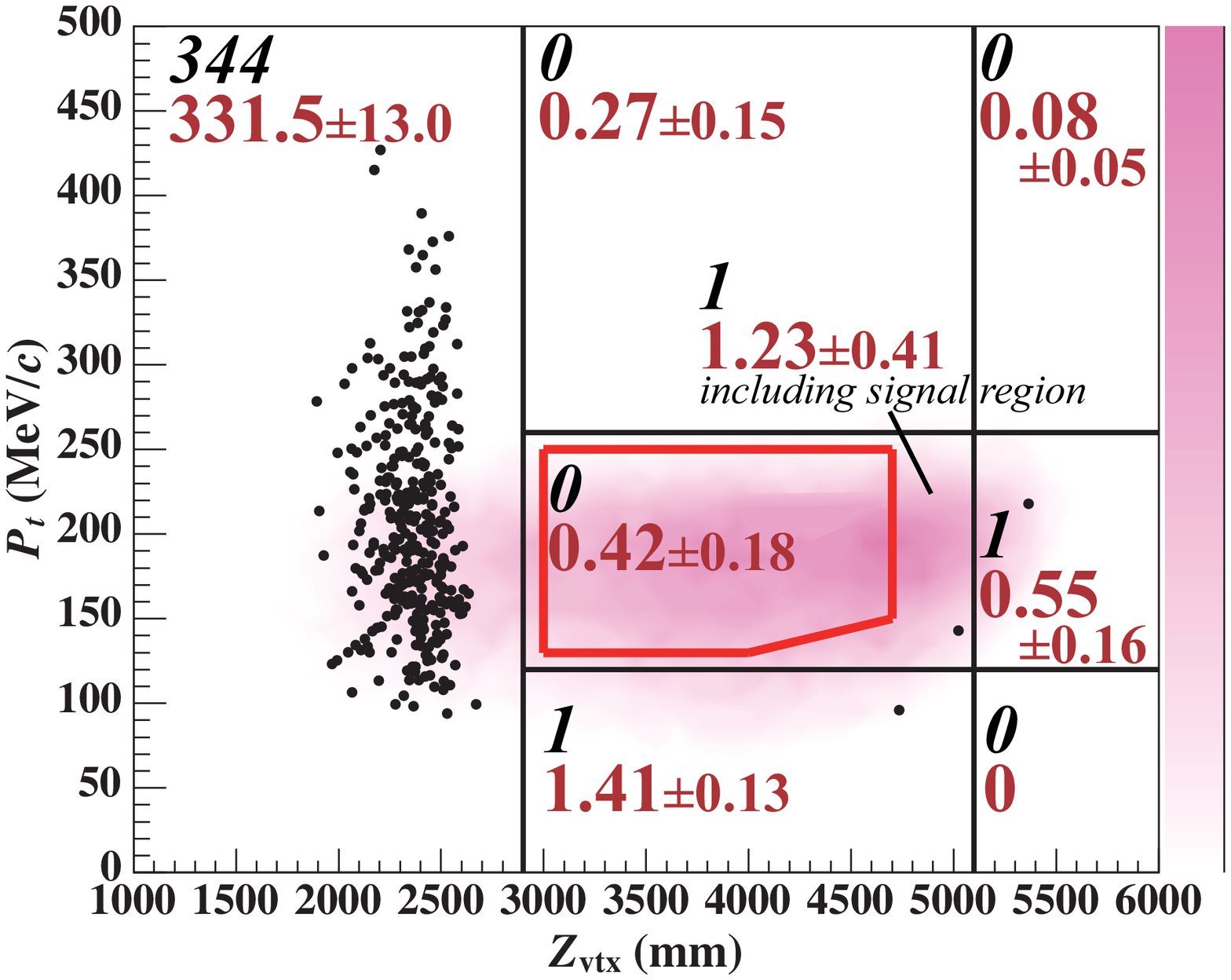}
			\caption{	Reconstructed $\pi^0$ transverse momentum ($P_t$) vs $\pi^0$ decay vertex position ($Z_{\rm vtx}$) plot of the events with all the cuts imposed.
					The region surrounded by red lines is the signal region. 
					The black dots represent observed events, and the contour indicates the $\klpionn$ signal distribution derived from the MC simulation.
					The black italic (red regular) numbers indicate the numbers of observed (expected background) events for the regions inside the lines.
					}
			\label{fig:FinalPtZ}
		\end{center}
	\end{figure}

\begin{figure}
		\begin{center}
			\includegraphics[width=1\linewidth]{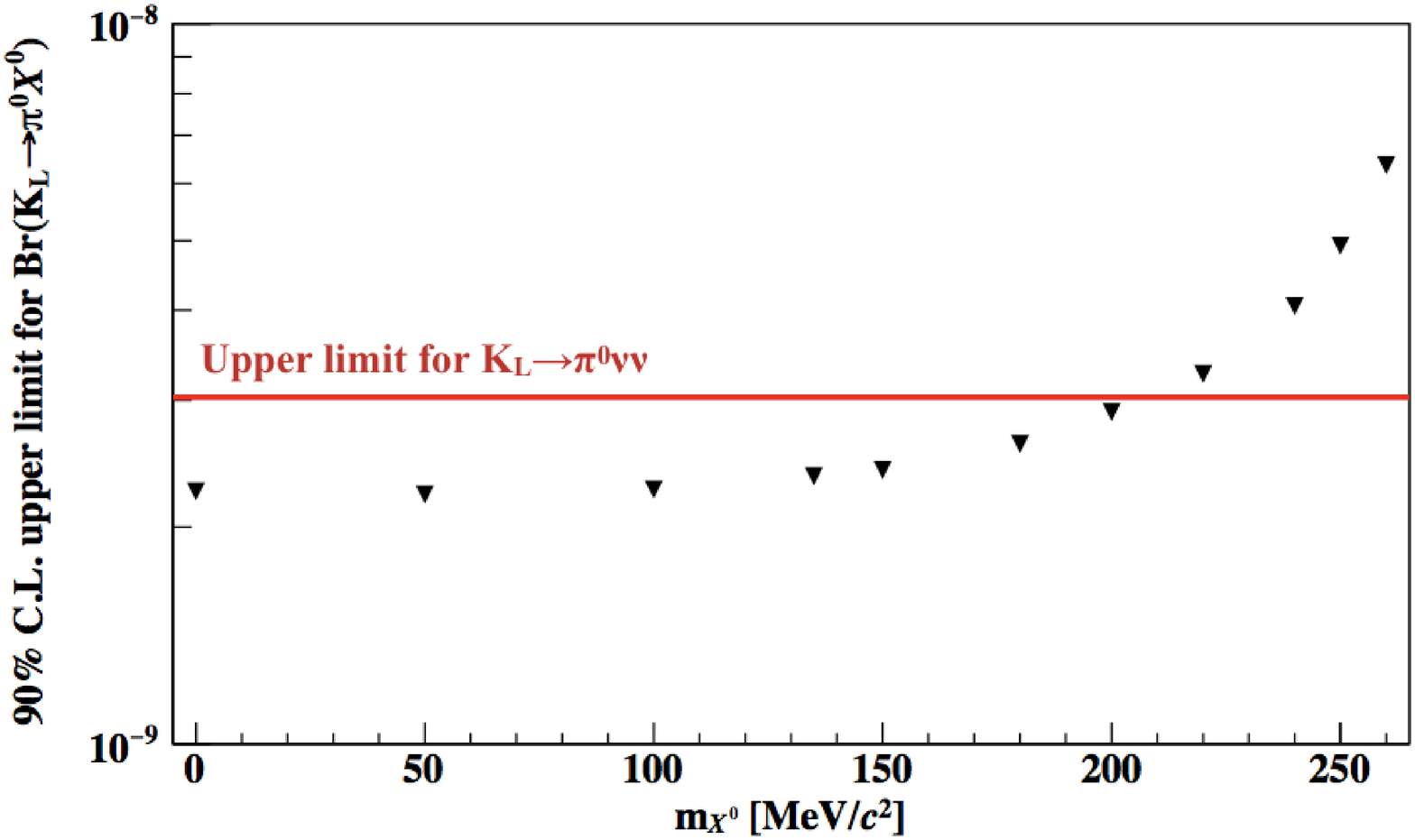}
			\caption{	Upper limit at the 90\% C.L. for the $\klpioxo$ branching fraction as a function of the $X^0$ mass. 
					For comparison, the limit for the $\klpionn$ decay is shown with the red line. 
					}
			\label{fig:Pi0X_UL}
		\end{center}
	\end{figure}

\paragraph*{Conclusions and prospects.}\hspace{-15pt}
	---After all the cuts were imposed, no signal candidate events were observed as shown in Fig.~\ref{fig:FinalPtZ}.
	Assuming Poisson statistics with uncertainties taken into account \cite{UpperLimit}, the upper limit for the branching fraction of the $\klpionn$ decay was
	obtained to be $3.0 \times 10^{-9}$ at the 90\% C.L.
	The upper limit for the $\klpioxo$ decay as a function of the $X^0$ mass ($m_{X^0}$) was also obtained as shown in Fig.~\ref{fig:Pi0X_UL};
	the limit for $m_{X^0} = m_{\pi^0}$ was set to be $2.4 \times 10^{-9}$ (90\% C.L.).
	These results improve the upper limit of the direct search by almost an order of magnitude.

	Based on this analysis, we developed necessary measures to reach better sensitivity.
	We anticipate to improve background rejection with data collected after 2015, 
	which corresponds to 1.4 times larger than the data in 2015,
	with a newly added veto counter in 2016 \cite{KAON2016_Togawa} and 
	more refined analysis methodologies, exploiting the substantially higher statistics of the collected control samples.
	
\begin{acknowledgments}	
	We would like to express our gratitude to all members of the J-PARC Accelerator and Hadron Experimental Facility groups for their support. 
	We also thank the KEK Computing Research Center for KEKCC and the National Institute of Information for SINET4. 
	We thank K. Yamamoto for useful discussions. 
	This material is based upon work supported by the Ministry of Education, Culture, Sports, Science, and Technology (MEXT) of Japan and 
	the Japan Society for the Promotion of Science (JSPS) under the MEXT KAKENHI Grant No.~JP18071006, 
	the JSPS KAKENHI Grants No.~JP23224007, No.~JP16H06343, and No.~JP17K05479, and 
	through the Japan-U.S. Cooperative Research Program in High Energy Physics; 
	the U.S. Department of Energy, Office of Science, Office of High Energy Physics, 
	under Awards No.~DE-SC{000}6497, No.~DE-SC{000}7859, and No.~DE-SC{000}9798; 
	the Ministry of Education and the National Science Council/Ministry of Science and Technology in Taiwan 
	under Grants No.~99-2112-M-002-014-MY3 and No.~102-2112-M-002-017; 
	and the National Research Foundation of Korea (2017R1A2B2011334 and 2017R1A2B4006359). 
	Some of the authors were supported by Grants-in-Aid for JSPS Fellows.
\end{acknowledgments}	
	
\bibliography{main.bbl}

\end{document}